**Consciousness Results when Communication Modifies the Form of Self-Estimated Fitness**

J. H. van Hateren

Johann Bernouilli Institute for Mathematics and Computer Science, University of Groningen, Groningen, The Netherlands

e-mail: j.h.van.hateren@rug.nl

**Abstract**  The origin and development of consciousness is poorly understood. Although it is clearly a naturalistic phenomenon evolved through Darwinian evolution, explaining it in terms of physicochemical, neural, or symbolic mechanisms remains elusive. Here I propose that two steps had to be taken in its evolution. First, living systems evolved an intrinsic goal-directedness by internalizing Darwinian fitness as a self-estimated fitness. The self-estimated fitness participates in a feedback loop that effectively produces intrinsic meaning in the organism. Second, animals with advanced nervous systems evolved a special form of communication that modifies the way each partner estimates fitness. The resulting change in intrinsic meaning is experienced subjectively as a primary form of consciousness. This primary form is subsequently used to generate, partly through internalized dialogue, more complex forms of consciousness, such as consciousness of the natural and social worlds, consciousness of the self, and language-dependent forms of consciousness.

**Keywords**  Self-estimated fitness · Evolution · Active causation · Intrinsic meaning · Dialogue · Subjective experience · Consciousness · Language

**Introduction**

Human behaviour is often accompanied by a subjective, conscious experience. There are strong indications that such experiences also occur, to varying degrees, in many animal species. The physical origin of this phenomenon is not understood: there is nothing in basic physics and chemistry that points to the existence of consciousness. It appears to have evolved fairly recently in evolution, because it presumably occurs only in animals with complex nervous systems.

    Here I will argue that its roots lie much earlier in evolution, as an extension of the basic Darwinian evolution scheme. The extension mixes deterministic and stochastic (random) forms of causation in a fitness-driven feedback loop. The result is an internalization of Darwinian fitness, leading to value and meaning intrinsic to the organism. The more recent development of animals with sophisticated nervous systems enabled a form of communication where this intrinsic meaning is, effectively, exchanged between animals. It is proposed here that the modification of intrinsic meaning, as required during such forms of communication, is accompanied by a subjective experience in the animal. The theory is extended, in a fairly straightforward way, to forms of consciousness that do not require communication between two animals, but occur within a single animal.

    The theory is strictly naturalistic, requiring only known physical components and mechanisms. The new properties, intrinsic meaning and consciousness, simply emerge from the rather special properties of the extended Darwinian evolutionary mechanism in combination with a particular mix of deterministic and stochastic causation. The article first explains the basic theory and its application to various forms of consciousness, then discusses the unity of consciousness that it implies, and finally discusses its relationship with several other theories of consciousness.

**Theory**

The theory presented here consists of a series of steps, some of which have been presented in more detail elsewhere. An overview and a computational analysis can be found in van Hateren (2014a) and an analysis of active causation as a fundamental property of life in van Hateren (2013). The consequences of the theory for agency and free will are analysed in van Hateren (2014c), which also includes a succinct discussion of consciousness. In order to keep the present article self-contained, I will first summarize the theory as it applies to communication between animals, and subsequently present the more detailed material with respect to consciousness.

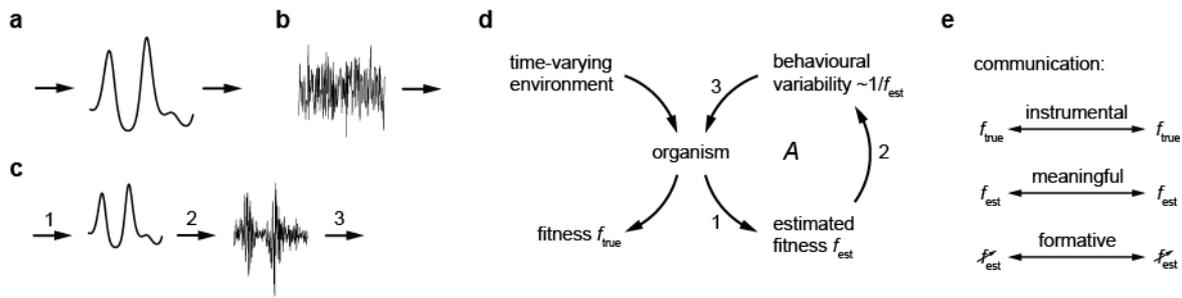

**Fig. 1** Origin of intrinsic meaning and consciousness. (**a**) Deterministic causation, where a system variable or property ('signal') is caused by (left arrow) or causes (right arrow) other signals; (**b**) stochastic causation, initiating new causal chains; (**c**) modulated stochastic causation, where a non-negative deterministic signal modulates the variance of a stochastic signal; (**d**) modulated stochastic causation (arrows 1-3 as corresponding to (**c**)) participates in a feedback loop $A$ where the organism uses an estimate $f_{est}$ of the true Darwinian fitness $f_{true}$ to generate behavioural variability. The result is behavioural freedom and an intrinsic goal-directedness in the organism, implying intrinsic meaning; (**e**) communication between two animals comes in three basic forms, either evaluated directly through $f_{true}$ (instrumental), indirectly through $f_{est}$ (meaningful), or aimed at changing the form of $f_{est}$ (formative). The resulting change in intrinsic meaning is conjectured to be experienced subjectively as consciousness by the animal

Causation and Self-Estimated Fitness

Figure 1 gives a schematic overview of the steps leading to an elementary form of subjective experience. The first point that needs to be appreciated is that there are different forms of causation at work in nature, as illustrated in Fig. 1a-c. In these diagrams, traces (called 'signals' below) symbolize system variables or properties, and arrows a causal relationship with earlier or later signals. The most basic form, deterministic causation (Fig. 1a), is also the standard form used in scientific explanations. Earlier signals cause a particular signal in a deterministic, predictable way, and this signal subsequently influences later signals similarly. The signal is thus part of a continuous, uninterrupted chain of causation, like a cog in a clockwork, which implies that it cannot start new causal chains. In contrast, stochastic causation (Fig. 1b), initiates new causal chains by definition, as it is not caused by (identifiable) earlier signals. Its statistical properties, like mean and variance, may be known, but they are constant in time and therefore cannot participate in causation. The unpredictable, individual fluctuations do the causal work. Stochastic causation can arise from thermal and quantum noise, or from disturbances coming from outside the system considered (systems are never completely isolated from their surroundings, not even from very remote parts, see e.g. Berry 1988). Although such noise may be microscopic in origin, it can easily be amplified to macroscopic scales if a system has components interacting in non-linear ways (similarly as in chaotic systems).

    A form of causation that is midway between deterministic and stochastic causation is illustrated in Fig. 1c. This so-called modulated stochastic causation is a key component of the theory presented here. It consists of two steps. In the first step, earlier signals cause a non-negative deterministic signal (left trace in Fig. 1c). This signal is subsequently used to modulate the variance of a stochastic signal (right trace), which then continues the causal chain (right-most arrow). The mean of the stochastic signal remains constant, and therefore does not participate in causation. The causal work is jointly done by a deterministic component (the varying variance) and a stochastic component (the individual fluctuations).

    Modulated stochastic causation plays an important role in genetic evolution, because it is a good model for how the stress response of cells modulates their mutation rate (high stress leads to high mutation rates and vice versa; Galhardo et al. 2007; see van Hateren 2014b for the consequences for evolutionary theory). However, it can also be extended to behavioural variability within the lifetime of individual organisms (van Hateren 2013; 2014a, c). I will concentrate on that here, because only the

behavioural timescale is relevant for consciousness. Figure 1d shows the basic mechanism. An organism is embedded in a time-varying environment, which includes other organisms. The Darwinian fitness of the organism is defined here, in its most basic form, as the expected number of offspring of the organism over its lifetime. It therefore depends on survival and reproduction. In a less basic form it includes fitness effects through related organisms (such as kin selection) and can involve organisms with shared interests (such as effectuated through social and cultural mechanisms). Note that fitness is defined here in a predictive, probabilistic sense, as a continuously updated expectation value, and it is therefore a function of time. It might fluctuate, for example declining when the organism gets seriously ill, and rising again if the organism recovers and flourishes thereafter.

Keeping fitness ($f_{true}$ in Fig. 1d) high is a prerequisite for Darwinian evolution. Previous selection must therefore have produced mechanisms in the organism that can be expected to promote fitness, at least on average. Such mechanisms may code directly for suitable properties and behaviours of the organism, or code for properties and strategies that allow it to learn during its lifetime and thereby acquire fitness promoting behaviours, again only on average. Either way, the form of causation is primarily deterministic. However, often the consequences of behaviour are not known in advance, where known is meant in a probabilistic way, as an expectation based on genetic memory established by natural selection in the past, or on physiological and neural memory established by the consequences of previous behaviour. With unknown consequences, the question is then how much behavioural variability is optimal. In van Hateren (2014a) it is argued that, similarly to the way mutation rates depend on cellular stress (which is inversely related to fitness), behavioural variability should also be driven by fitness. When current fitness is high, behaviour is presumably adequate, and it should not change much (low variability), just a little to allow for the possibility to find a behaviour that is even better. When current fitness is low, behaviour is presumably inadequate, and large changes in behaviour (high variability) are necessary in order to avoid the (potentially lethal) consequences of continued low fitness. Behaviour needs to be varied until a behaviour is found that produces higher fitness, upon which variability can be lowered. Model calculations (van Hateren 2014a) indicate that this is indeed an evolvable mechanism.

A problem with this scheme is that fitness itself, $f_{true}$, is not available to the organism. It cannot be observed directly, and to determine it would require a detailed simulation of the organism and its interaction with its environment, including other organisms. Such a simulation is clearly out of reach for the organism. However, what it can do is to make an approximate estimate[1] of its own fitness, $f_{est}$, and use that to drive its behavioural variability. The estimate can make use of a large range of indicators the organism could have about itself, such as its physiological state, and its environmental conditions. For example, low fitness associated with an internal lack of nutrients can be inferred from internal state variables, and the presence or absence of nutrients in the environment can be assessed through the organism's senses. The self-made fitness estimate is assumed to be present only implicitly, represented in a distributed way throughout the organism's physiology and nervous system. The way it affects behaviour will be similarly diffuse. Although $f_{est}$ can only approximate $f_{true}$, the better the approximation, the better it can increase fitness. The estimate is therefore under Darwinian selection pressure to be as close to $f_{true}$ as possible given the constraints and means of the organism.

The loop marked *A* in Fig. 1d shows this scheme. It is a feedback loop (i.e., with cyclical causation), running through the organism. A continuously updated $f_{est}$ drives behavioural variability, with variability roughly inversely related to $f_{est}$. This relationship is symbolized by $\sim 1/f_{est}$, although its actual form may be more complicated and is also subject to Darwinian selection. The causation in this loop conforms to the modulated stochastic causation of Fig. 1c (see corresponding arrows marked 1-3; strictly speaking the left trace in Fig. 1c corresponds to $1/f_{est}$, not to $f_{est}$). Only the variance of the behaviour should be modulated, not the mean, because the part of the behaviour driven by this process is the part for which the fitness consequences are not known. Behavioural change should therefore be undirected, i.e., not into one particular direction rather than another (on average). The directed part of behavioural change is under control of the genetic and learned mechanisms mentioned above, which are primarily deterministic.

---

[1] The term 'estimate' is used here in the theoretical, technical sense as in estimation theory. It just refers to a variable that approximates another variable. Note that it is devoid of any deliberateness by the organism, it is just present in its physiology. Also note that it is in no way related to the estimates that a scientist might want to make of the true fitness. The latter is part of doing science, whereas the organism's estimate is part of its biological functioning, independent of whether there are scientists or not.

Active Causation and Intrinsic Meaning

The fact that modulated stochastic causation takes part in a feedback loop that is indirectly driven by Darwinian fitness (because $f_{est}$ estimates $f_{true}$) has two important consequences (van Hateren 2014a). First, it produces a form of causation for which I have coined the term active causation. The feedback loop produces, each time it is traversed, a behavioural change that is mostly, but not completely stochastic. The tiny part that is not stochastic is due to the fact that the variance is driven by $f_{est}$. However, this tiny part gradually accumulates, with each time the loop is traversed, into a behavioural trajectory (the sequence of subsequent behaviours) that gradually becomes, statistically, strongly dependent on $f_{est}$ (van Hateren 2014a). The resulting behavioural trajectory is mostly stochastic in its details, but as a whole it can only be understood by the action of $f_{est}$. In other words, the behaviour as it manifests itself on a longer timescale (many loopings through $A$) is neither completely deterministic nor completely stochastic, but driven by the organism itself (through its estimate of its own fitness). It is therefore an active form of causation, presumably unique to living systems (van Hateren 2013). Effectively, it provides the organism with some behavioural freedom, i.e., an elementary form of agency (see also Heisenberg 2009).

    A second consequence of the feedback loop is that it produces a genuine goal in the organism, namely high $f_{est}$. Whereas high $f_{true}$ cannot be seen as a genuine goal of the organism, because it is merely a consequence of an external physicochemical process (selection against low $f_{true}$ in the past), this is different for $f_{est}$. The form of $f_{est}$ is not fixed, as long as its value is sufficiently close to $f_{true}$ such that the mechanism using $f_{est}$ is evolvable and evolutionary stable. The feedback loop produces behavioural freedom and agency, replacing the deterministic causation of processes involving $f_{true}$ by the active causation of processes involving $f_{est}$. High $f_{est}$ should therefore be viewed as a genuine goal of the organism. The way by which the organism evaluates $f_{est}$ signifies which internal and external parameters the organism estimates to be important for its own fitness. I have therefore coined the term intrinsic meaning for the specific form of the $f_{est}$ of a particular organism (van Hateren 2014a). Meaning is used here in the general sense of import, significance, value, and purpose, i.e., similar to its use in 'the meaning of an action' instead of its use in 'the meaning of a word'. Intrinsic meaning should be seen as not just a theoretical construct, but as a genuine physical property, evolved through Darwinian evolution and only present in living systems. It is, in principle, not different from other genuine physical properties that only occur in material systems with very special (dynamic) structures, such as the property of high-temperature superconductivity in specific materials, or the property of some spherical objects that they can roll on a plane.

    The scheme of Fig. 1d is easiest to understand when $f_{est}$ has a simple, one-dimensional form, where it drives a single behaviour and is evaluated in a simple way from the state of the environment and the properties of the organism (see e.g. the computational model in van Hateren 2014a). In more realistic cases, $f_{est}$ would have a range of different inputs, and it would need to drive the variability of a range of different behaviours. Then the partial fitness effects of each input and each behavioural output would need to be taken into account and properly weighted. This will quickly become intractable in realistic cases, where the form of $f_{est}$ is expected to become highly complicated (with complex dynamics involving nonlinearities and memory) and the number of inputs and outputs large and interdependent. The current theory should therefore be primarily seen as a conceptual aid in understanding the phenomena of meaning and consciousness, and not so much as a first step towards a comprehensive quantitative model.

Formative Communication and Consciousness

The fact that animal behaviour is evaluated through two forms of fitness, passively through $f_{true}$ and actively through $f_{est}$, implies that there are three basic forms of communication possible between animals (van Hateren 2014d). The simplest combinations of these forms are illustrated in Fig. 1e (although in practice different forms will usually mix with different weights and in different combinations for each partner in the communication). Communication is called instrumental when it only involves $f_{true}$. It is deterministic and reflex-like, and does not involve behavioural freedom. Communication is called meaningful when it involves $f_{est}$. This produces some behavioural freedom, and communication is evaluated depending on the forms of $f_{est}$, i.e., depending on the intrinsic meaning of the communication to each of the partners. In contrast to $f_{true}$, $f_{est}$ is intrinsic to each animal, and it is therefore in principle under its control and modifiable. This implies that there is a third form of communication possible, called formative. In formative communication the very form of $f_{est}$ is modified, a modifiability symbolized by the oblique arrow through $f_{est}$ in the lower diagram of Fig. 1e. The basic idea is that changing the form of $f_{est}$ can yield fitness benefits for both partners in the communication, for example by forming mother-infant bonds in mammals, and partner bonds in mammals and birds. It works best when there is a stable incentive for cooperation, because modifying the form of $f_{est}$ is inherently a rather risky operation. It can be abused easily by other organisms, and it will endanger the organism when $f_{est}$ strays too far from $f_{true}$. Presumably, it can therefore only operate in an evolutionary stable way by using a range of checks and balances that require a sufficiently advanced nervous system.

The main conjecture made in this article is the following. The form of $f_{est}$ embodies the intrinsic meaning of an animal, the value it assigns to its internal and external state. Intrinsic meaning is a genuine physical property, only present in living systems. By modifying the form of $f_{est}$ during acts of communication, a genuine physical property is thus modified. It is known from physics that changing one physical entity may produce another, qualitatively different physical entity. For example, a changing electric field produces a magnetic field. The conjecture is, then, that modifying intrinsic meaning originates another genuine physical phenomenon, consciousness. Thus what is experienced subjectively is associated with the change in intrinsic meaning. Although intrinsic meaning is normally an implicit, distributed property of the organism, parts of it need to be made explicit during acts of formative communication. The reason is that there is no way intrinsic meaning can be transferred directly from one organism to the other: it is a physical phenomenon that only exists within organisms. The only way to exchange intrinsic meaning between communicating partners is therefore indirectly, via regular physical acts of behaviour. The distributed, implicit meaning needs to be extracted and made explicit by the communicator, and coded suitably in physical behaviour. The recipient needs to perceive this physical behaviour, and translate it into a form that can be assimilated into its own, distributed intrinsic meaning. Presumably, subjective experience accompanies the acts of extracting and assimilating intrinsic meaning, which in either case is expected to be modified by the very act.

Although there are many ways by which intrinsic meaning may be modified through formative communication, it is assumed here that its evolutionary origin lies in pair-bonding, such as the mother-infant bond in mammals. From this basis, more complex forms of consciousness can be constructed during development, as detailed in the next section.

Origins of Various Forms of Consciousness

Pair-bonding seems to be the most plausible evolutionary and developmental origin of the subjective experience associated with formative communication. However, whether this is the case is ultimately an empirical question that would need to be assessed in future research. The ways more complex forms of consciousness can be constructed from the basic one as described in this section should therefore primarily be seen as draft proposals. The order and specific paths presented here may be different and more complex in actual organisms: evolution needs to work from existing material and incrementally, and thus adds layer upon layer. What it produces is often sophisticated and tortuous at the same time. Furthermore, many of the mechanisms, in particular the ones involving symbolic communication, will have a strong cultural component, adding to the complexity.

Figure 2a shows at (1) the basic pair-bond in a symbolic way. The double-headed arrow stands for a dialogue (two-way, continued formative communication) between two subjects, $S_1$ and $S_2$. As the prototypical pair-bond I will take the human mother-infant bond, which has been particularly well studied (e.g., Trevarthen and Aitken 2001; Reddy 2003). In all of Fig. 2 (except for the language-related diagrams in Fig. 2e), $S_1$ stands for the infant and $S_2$ for the adult. The basic bond at (1) produces a simple form of consciousness in $S_1$, a subjective experience of $S_2$ in relation to itself. At (2) the basic bond is used to gradually acquire an internalized version of $S_2$, symbolized by $\hat{S}_2$. This internal model of $S_2$ then becomes sufficiently realistic to maintain a (partly simulated) dialogue between $S_1$ and $\hat{S}_2$, thus enabling consciousness of $S_2$ without the presence of $S_2$ (3).

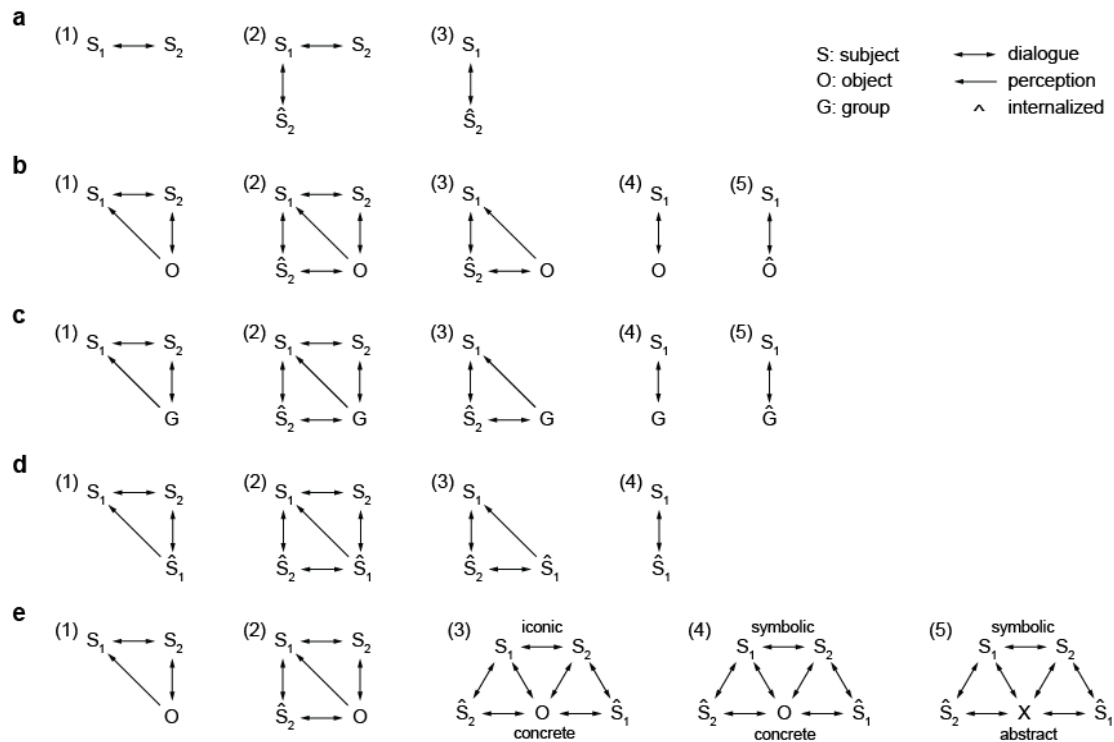

**Fig. 2** Origin of various forms of consciousness. (**a**) Basic mother-infant bond (1), with the double arrow a dialogue (continued formative communication, implying subjective experience) between infant $S_1$ and adult $S_2$. $S_1$ gradually (2) internalizes $S_2$ ($\hat{S}_2$), and uses an internalized dialogue with $\hat{S}_2$ as a source of consciousness even when $S_2$ is absent (3); (**b**) consciousness of the natural world arises when objects O are first perceived in interaction with $S_2$ (1), and $S_2$ and the interaction are internalized (2), even when $S_2$ is absent (3); this finally results in simulated dialogue with O (4) and its internalized version (5); (**c**) similarly as (**b**) for a group G of individuals, leading to consciousness of the social world; (**d**) similarly as (**b**) for the self as experienced as an object of attention to $S_2$, eventually leading to an elementary form of self-consciousness; (**e**) the first two steps are identical to (**b**), but in step (3) the direct dialogue $S_1$-$S_2$, and the dialogues $S_1$-$\hat{S}_2$ and $S_2$-$\hat{S}_1$ are retained, leading to enhanced communication, first iconic (3) and symbolic (4) dialogues about concrete objects, and finally (5) symbolic dialogues about abstract topics X. Scheme (5) without the presence of $S_2$ implies symbolic thought (internalized dialogue between $S_1$ and either $\hat{S}_1$ or $\hat{S}_2$ about X)

Although the basic pair-bond is used here to derive more complex forms of consciousness below, it implies an even simpler form of subjective experience. Subjective experience is assumed to accompany any act of formative communication, either as sender or recipient. A dialogue is not strictly necessary, although it is expected to sustain and, through positive feedback, amplify consciousness. Even if $S_2$ is absent, a new-born infant crying in response to a painful stimulus will experience that subjectively according to the present theory, because it is an act of formative communication. It slightly modifies the infant's intrinsic meaning, even if the attempted dialogue fails.

The natural world can become part of subjective experience as shown in Fig. 2b. The basic $S_1$-$S_2$ bond is extended with an interaction of $S_2$ with an object O. This is in the form of a (simulated) dialogue, because $S_2$, the adult, has already formed an internalized model of O. Initially, $S_1$ lacks such an internal model, and perceives O without subjective experience (1). However, the internalized version of $S_2$ as formed according to Fig. 2a, $\hat{S}_2$, can be gradually extended with the interaction with O (and its implicit version $\hat{O}$ as used by $S_2$), as shown in (2). Once this has become sufficiently realistic, $S_2$ need not be present any more (3). Finally, the intermediate $\hat{S}_2$ can fade away, and $S_1$ directly interacts with O (4), using an internal model $\hat{O}$, or even interacts with this model without the presence of O (5). In either case, the interaction is a simulated dialogue, with a modifying intrinsic meaning accompanied by subjective experience.

The social world, symbolized by G, a group of individuals, can become internalized in a similar way as the natural world (Fig. 2c). An obvious difference with O is that G, or at least some members of G, can engage in a genuine dialogue with $S_1$. This makes the interactions and internalized model more complex than in the case of the natural world (although the presence of other living and perhaps conscious species might complicate the latter case as well).

Establishing consciousness of the self may proceed according to Fig. 2d. In the dialogue between $S_1$ and $S_2$, the latter will implicitly use an internalized version of $S_1$. Perceiving this is more difficult for $S_1$ than perceiving O as in Fig. 2b. Nevertheless, the behaviour of $S_2$ when paying attention to $S_1$ bears resemblances to when $S_2$ pays attention to O. $S_1$ may perceive this (1) as being the object of attention (Reddy 2003). Once $S_1$ has developed a realistic model of $S_2$, such a model may get extended, by inference, with an internalized model of $S_1$, $\hat{S}_1$, as apparently used implicitly by $S_2$ (2). Once this is sufficiently realistic for a simulated dialogue, $S_2$ needs not be present any more (3). Finally, a direct simulated dialogue between $S_1$ and $\hat{S}_1$ (4) produces an elementary form of self-consciousness. Note that self-consciousness as conjectured here does not involve self-referentiality (i.e., there is no circularity): $S_1$ and $\hat{S}_1$ are two different entities, with the subjective experience produced, as before, by the formative change in the $f_{est}$ of $S_1$ while communicating with the simulation (model) $\hat{S}_1$. Obviously, this change may subsequently lead to an adjustment of $\hat{S}_1$, but this is then just the regular cyclicism of feedback, rather than circularity.

Symbolic communication may arise as depicted in Fig. 2e. The first two steps are identical to Fig. 2b, establishing a basic dialogic connection with an object O, using an internalized $\hat{S}_2$ as intermediary. In contrast to Fig. 2b, the dialogues with $S_2$ and $\hat{S}_2$ are maintained even after the possibility of a dialogue with O (implicitly using $\hat{O}$) has been established. This leads to scheme (3), where for completeness the implicit $\hat{S}_1$ used by $S_2$ has been added. This scheme greatly facilitates dialogue about O between $S_1$ and $S_2$, because both maintain internalized versions of each other. In effect, $S_1$ can communicate taking $S_2$'s perspective into account, and vice versa. The signs used for communication may initially be similar (iconic) to concrete objects (3), but gradually become symbolic (4), and eventually refer to abstract objects (X, such as categories, social events, and ideas) as well (5). Symbolic communication, particularly in the form of language, is a specialization of humans (Deacon 1997), and presumably requires evolved motivations, such as a propensity for sharing intentionality (Tomasello and Carpenter 2007) and a willingness to cooperate (Richerson and Boyd 2005).

A final stage for $S_1$ is to use the scheme of Fig. 2e without the presence of $S_2$. Conscious, symbolic thought then involves an internalized dialogue of $S_1$ with either $\hat{S}_1$ or $\hat{S}_2$ about X. Symbolic communication and thought will subsequently enhance consciousness of the other, the world, and the self, as complex extensions to the schemes of Fig. 2a-d.

The Unity of Subjective Experience

In the previous paragraph several forms of consciousness were discussed, and even a single form, say consciousness of the natural world, usually contains a range of components. The question is, then, how there can be unity of consciousness. Figure 3 explains why the theory presented here must lead to such a unity. The organism has only a single $f_{true}$, and therefore also a single $f_{est}$ (barring pathology), and only this single $f_{est}$ can be modified in formative communication. It is this fluidity of intrinsic meaning that gives rise to consciousness. There may be many components contributing to $f_{est}$ (the time-varying states in Fig. 3 that are used by the organism to assess intrinsic meaning), and there may be many behavioural variabilities under control of $f_{est}$, and there may be many ways the organism can engage in dialogue, all subjectively experienced (as different qualia), but there is only one $f_{est}$. If all is well, there must be unity of consciousness.

How the unity of consciousness is realized in the organism's physiology and nervous system is likely to be opaque. There is no reason for $f_{est}$ to be localized, as all parts of the organism can, in principle, play a role in the system that embodies $f_{est}$. Obviously, if $f_{est}$ is to be a good estimate of $f_{true}$, it needs to be well coordinated across the organism. However, such a coordination serves the organism's fitness, and should not be seen as a means to ensure unity of consciousness. The coordination is interesting from a scientific point of view in that it provides information on how the organism is realizing its $f_{est}$, i.e., its intrinsic meaning. But the interpretation would be extraordinarily difficult, because fitness itself is extremely complex for a social and cultural species such as the human one. For the same reason it may be similarly challenging to integrate neural correlates of consciousness (reviewed in Tononi and Koch 2008) into a coherent picture that could be interpreted within the present framework, i.e., as implementations of $f_{est}$, the *A* loop of Fig. 1d, and external and internalized dialogues.

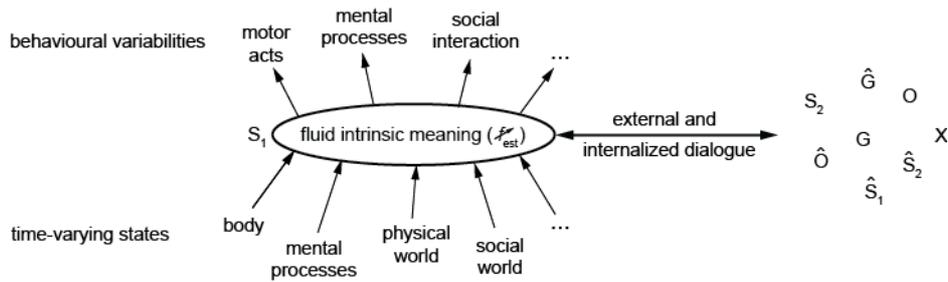

**Fig. 3** Unity of subjective experience. Organisms have only one true fitness and therefore should have only one estimated fitness $f_{est}$, and modifying this one $f_{est}$ therefore must lead to a unitary experience. Nevertheless, the form of $f_{est}$ is expected to be embodied in a distributed way, and to be highly complex, with many state variables (of body, mind, and environment) contributing, with the variability of many different behaviours controlled, and with consciousness arising from a variety of internalized and external dialogues (leading to qualitatively different conscious experiences, the qualia)

**Discussion**

In this article I have argued that consciousness arises from acts of formative communication, usually during dialogue. Formative communication is a special form of communication that modifies the intrinsic meaning as embodied in animals in the form of a self-estimated fitness. Such a fitness participates in a feedback loop that finds a middle ground between determinism and stochasticity, and equips the animal with behavioural freedom in addition to intrinsic meaning. Dialogue and elementary consciousness are assumed to originate from a basic bond, such as the mammalian mother-infant bond. This bond is subsequently used for establishing more complex forms of consciousness, such as consciousness about the natural world, about the social world, self-consciousness, symbolic dialogue with others about concrete or abstract objects (language), and internalized symbolic dialogue (thought).

    The theory presented here is consistent with Darwinian evolution, although it extends the basic scheme using only extrinsic fitness ($f_{true}$) with one where an internalized, self-estimated fitness ($f_{est}$) plays a role as well. The consequence is a scheme that allows for agency, meaning, and consciousness (van Hateren 2014a, c). Fitness implies embodiedness (e.g., Damasio and Carvalho 2013), because without a body there would be no survival, reproduction, and death. The present theory is therefore rather different from purely symbolic and informational approaches to understanding meaning and consciousness (see also Searle 2013). However, the present theory also implies that embodiedness as such is not sufficient for generating meaning and consciousness: embodiedness without the *A* loop of Fig. 1d would not produce active causation and intrinsic meaning, and therefore also no consciousness in the formative communication of Fig. 1e.

    Similarly, fitness implies embeddedness and enaction (e.g., Thompson 2007; McGann et al. 2013; Engel et al. 2013), as it is determined to a large extent by how the organism interacts with its environment. This also follows from the way conscious perception of the world is understood here, as in Fig. 2b. Formative communication with O or its internalized version implies modifying $f_{est}$. It is therefore action-oriented, because Darwinian fitness is ultimately determined by the consequences of action and interaction. But also embeddedness and enaction are not sufficient for generating meaning and consciousness, for the same reasons as stated above. Theories stressing embodiedness and embeddedness often derive a concept of value from elementary life-sustaining processes, such as homeostasis and metabolism. However, this would not work without the value-producing *A* loop of Fig. 1d, because the basic Darwinian loop (using only $f_{true}$) is inherently value-free, where life and death are just consequences of the process, not related to values and goals (see also Davies 2009, pp. 86-87, and van Hateren 2014a, c).

A recent theory (Tononi 2008; Edlund et al. 2011) proposes to explain consciousness as an inherent property of integrated information, i.e., the excess information in the system produced by the integration of its parts over the total of the individual parts. The theory presented here presumably implies that a conscious organism would score high on integrated information, because $f_{est}$ is unitary and distributed at the same time (as illustrated in Fig. 3). Complex forms of consciousness, such as involving symbolism (Fig. 2e), would require well-connected, but complex subsystems. However, the reverse is not true. A system which scores high on integrated information, but without the *A* loop of Fig. 1d, would completely lack meaning and consciousness according to the present theory.

From the present perspective it is clear that, although consciousness is a phenomenon that has evolved through Darwinian evolution, the question what fitness benefits it would yield is slightly ill-posed. The fitness benefits come primarily from formative communication, enabling a flexible adjustment of the self-estimated fitness of the communicative partners. This may facilitate cooperation to such an extent that it presumably produces higher fitness in each partner than they would have had without the dialogue, at least on average. Subjective experience is then just the phenomenon that accompanies a modifying intrinsic meaning. Both intrinsic meaning and consciousness are presumably unique to life, the former a property of all forms of life (van Hateren 2013), and the latter limited to specific species with a sufficiently social lifestyle and a sufficiently advanced nervous system. Nevertheless, advanced forms of consciousness may indeed produce fitness benefits by themselves. The strong reliance on internalized forms of dialogue, particularly in humans, suggests such additional fitness benefits, for example by enabling more flexible and deliberate forms of engagement with the physical and social worlds (van Hateren 2014c). The high levels of formative communication found in humans, as in language and symbolic thought, are accompanied by correspondingly high levels of consciousness alongside extraordinary fitness.